\documentclass[pre,12pt,preprint,superscriptaddress, amsmath, amssymb, floatfix]{revtex4-1} 

\renewcommand{\thetable}{\arabic{table}}
\renewcommand{\thefigure}{\arabic{figure}}

\usepackage{graphicx}  
\usepackage{graphics}
\usepackage{dcolumn}   
\usepackage{bm}        
\usepackage{amssymb}   
\usepackage{color}
\usepackage{amsmath}
\usepackage{threeparttable}
\usepackage[font=scriptsize]{caption}
\usepackage{url}
\usepackage{makecell}
\usepackage{diagbox}
\usepackage{booktabs}
\usepackage[figuresright]{rotating}
\usepackage{changes}
\usepackage{hyperref}

\definecolor{darkred}{RGB}{139,0,0}
\definecolor{chartreuse}{RGB}{127,255,0}
\definecolor{goldenrod}{RGB}{218,165,32}
\definecolor{gray}{RGB}{127,127,127}

\definecolor{Magenta}{RGB}{255, 0,255}
\definecolor{Orange}{RGB}{255,165, 0}
\definecolor{Gray}{RGB}{127,127,127}

\begin{document}

\title{Accelerated functional brain aging in major depressive disorder: evidence from a large scale fMRI analysis of Chinese participants}
\author{Yunsong Luo}
\affiliation{College of Computer and Information Science, Southwest University, Chongqing, 400715, P. R. China}
\author{Wenyu Chen}
\affiliation{College of Computer and Information Science, Southwest University, Chongqing, 400715, P. R. China}
\author{Jiang Qiu}

\affiliation{Key Laboratory of Cognition and Personality(SWU), Ministry of Education, Chongqing, 400715, P. R. China}
            
\affiliation{School of Psychology, Southwest University (SWU), Chongqing,400715, P. R. China}
            
\affiliation{Southwest University Branch, Collaborative Innovation Center of Assessment Toward Basic Education Quality at Beijing Normal University,Chongqing, 400715, P. R. China}  
\author{Tao Jia}
\email{tjia@swu.edu.cn}
\affiliation{College of Computer and Information Science, Southwest University, Chongqing, 400715, P. R. China}

\date{\today}

\begin{abstract}
Major depressive disorder (MDD) is one of the most common mental health conditions that has been intensively investigated for its association with brain atrophy and mortality. Recent studies reveal that the deviation between the predicted and the chronological age can be a marker of accelerated brain aging to characterize MDD. However, current conclusions are usually drawn based on structural MRI information collected from Caucasian participants. The universality of this biomarker needs to be further validated by subjects with different ethnic/racial backgrounds and by different types of data. Here we make use of the REST-meta-MDD, a large scale resting-state fMRI dataset collected from multiple cohort participants in China. We develop a stacking machine learning model based on 1101 healthy controls, which estimates a subject's chronological age from fMRI with promising accuracy. The trained model is then applied to 1276 MDD patients from 24 sites. We observe that MDD patients exhibit a $+4.43$ years ($\text{$p$} < 0.0001$, $\text{Cohen's $d$} = 0.35$, $\text{95\% CI}:1.86 - 3.91$) higher brain-predicted age difference (brain-PAD) compared to controls. In the MDD subgroup, we observe a statistically significant $+2.09$ years ($\text{$p$} < 0.05$, $\text{Cohen's $d$} = 0.134483$) brain-PAD in antidepressant users compared to medication-free patients. The statistical relationship observed is further checked by three different machine learning algorithms. The positive brain-PAD observed in participants in China confirms the presence of accelerated brain aging in MDD patients. The utilization of functional brain connectivity for age estimation verifies existing findings from a new dimension.
\end{abstract}

\maketitle

\section{Introduction}

Global population aging is expected to be one of the prominent social changes of the 21st century \cite{vos2012years}. The resulting burden of age-related functional decline and disease would challenge all sectors of society, especially healthcare \cite{dinsdale2021learning}. Therefore, understanding the biological link between aging and disease risk becomes increasingly important to provide effective care and treatment \cite{cole2017predicting}. Aging can be regarded as a dynamic process in which an individual gradually losses her function as cumulative age-related damage accumulates. The brain structure and function are also significantly changed during this process \cite{harman2001aging}. As the central nervous system may age dissimilarly to the rest of the body \cite{isaev2018accelerated}, brain-specific aging markers may be of particular importance in assessing the risk of cognitive decline and propensity to neurodegenerative diseases \cite{dafflon2020automated,buckley2020age}.

Accelerated aging of the brain refers to the phenomenon that an individual's brain appears older compared with the expected chronological age. Brain predicted age difference (brain-PAD), calculated as the difference between the estimated brain age from neuroimaging and the chronological age, is predisposed to be associated with the risk of cognitive aging or age-related brain disorders. This neuroimaging-based biomarker is observed in several neurological disorders \cite{kuo2020large,mishra2021review}, including schizophrenia \cite{shahab2019brain,koutsouleris2014accelerated,schnack2016accelerated}, Alzheimer's disease \cite{gonneaud2021accelerated,moradi2015machine,iturria2016early}, epilepsy \cite{sone2021neuroimaging}, multiple sclerosis \cite{cole2020longitudinal}, and traumatic brain injury \cite{cole2015prediction}. Furthermore, an association between brain-PAD and mortality \cite{cole2018brain} has also been reported.

Recent research set out to explore the relationship between accelerated brain aging and major depressive disorder (MDD), a widespread, debilitating, and disabling psychiatric disorder \cite{ferrari2013burden} associated with cellular senescence \cite{verhoeven2018depression,verhoeven2014major} and cognitive decline \cite{john2019affective}. Despite the positive association reported \cite{koutsouleris2014accelerated,han2021stage,christman2020accelerated,kaufmann2019common,dunlop2021accelerated,drobinin2021developmental}, current studies still present several limitations. First, the size of samples can have a great influence on the stability of the relationship discovered. While subjects with diverse sizes are investigated \cite{besteher2019machine,kuo2020large}, only one study analyzes data from more than 1,000 individuals \cite{han2020brain}, to the best of our knowledge. Moreover, most of the current studies are based on Caucasian subjects. In the meanwhile, the brain age is usually estimated from the structural MRI, with the grey or white matter volume and cortical thickness as the key feature. The generalizability of the association needs to be further verified in subjects from different ethnic and cultural backgrounds and tested by other types of neuroimaging data. Finally, previous studies often rely on only one machine learning algorithm to estimate brain age. Since different algorithms yield different estimations, it is reasonable to suspect that the conclusion drawn is algorithm sensitive. The statistically significant association originally reported may vanish when a new algorithm is applied.

To cope with these limitations, we make use of the resting-state functional magnetic resonance imaging (rsfMRI) data \cite{vergun2013characterizing} collected by the REST-meta-MDD \cite{yan2019reduced,yang2021disrupted}, which is a coordinated multisite project from China containing over 1000 MDD patients and normal controls. We utilize three different machine learning algorithms to estimate brain age from resting-state functional connectivity \cite{liem2017predicting,zhai2019predicting,li2018brain}. We further propose a stacking model to combine results from the three algorithms to reach a more optimal age estimation. We conduct separate analyses on results obtained from each algorithm to check the robustness of the conclusion drawn. We confirm the existence of the positive association between accelerated brain aging and MDD based on subjects in China. The brain-PAD is significantly higher in MDD patients compared to controls and the conclusion is not affected by the machine learning algorithm applied. We separately analyze MDD patients with different depression severity, illness duration, episode status, and medication status to investigate the association between brain-PAD with demographic (age, sex) and clinical characteristics. We find a significant correlation between brain-PAD and illness duration in MDD patients as well as a higher brain-PAD in antidepressant users than in medication-free patients.

\section{Methods}
\subsection{Samples}
We conduct this study through rsfMRI indices of MDD patients and matched controls (aged 12-82 years) from the REST-meta-MDD consortium, which consists of 25 research groups from 17 hospitals in China. All MDD patients are hospital diagnosed and conducted at least a T1-weighted structural scan and a rsfMRI scan. All subjects agree to provide diagnosis, age, gender, and education years. The 17-item Hamilton Depression Rating Scale (HDRS) is administered to some patients with other tabular data provided, including episode status (if the patient's prior and current episodes are diagnosed as MDD according to ICD10 or DSM-IV), medication status (whether antidepressants are used), and illness duration. After quality control, we reach a sample set of 1276 MDD patients and 1101 controls from 24 sites. Additional details regarding inclusion and exclusion criteria, demographic measures, and medication status are provided in Supplementary Table \ref{Samples of selected sites} and Table \ref{Data acquisition parameters of selected sites}. Written informed consent is signed by participants at each local site, and all data are de-identified and anonymized. Besides, approvals from the local institutional review board and ethics committee are granted at all sites.

\subsection{rsfMRI data preprocessing and functional brain network construction}
The Data Processing Assistant for Resting State fMRI (DPARSF) \cite{yan2010dparsf} is used as a standardized preprocessing pipeline. To obtain the functional connectivity, we first extract 116 averaged blood oxygen level-dependent (BOLD) signals based on the Automated Anatomical Labeling (AAL) atlas. Next, we calculate the Pearson correlation coefficients between the BOLD activity time series. Fisher’s r-to-z transform is applied to get the whole-brain functional network from the functional connectivity matrix of each subject. More details of the data preprocessing process are presented in Supplementary Information S2.

\begin{figure}[htb!]
    \centering
    \includegraphics[width=1\textwidth,height=0.6\textheight]{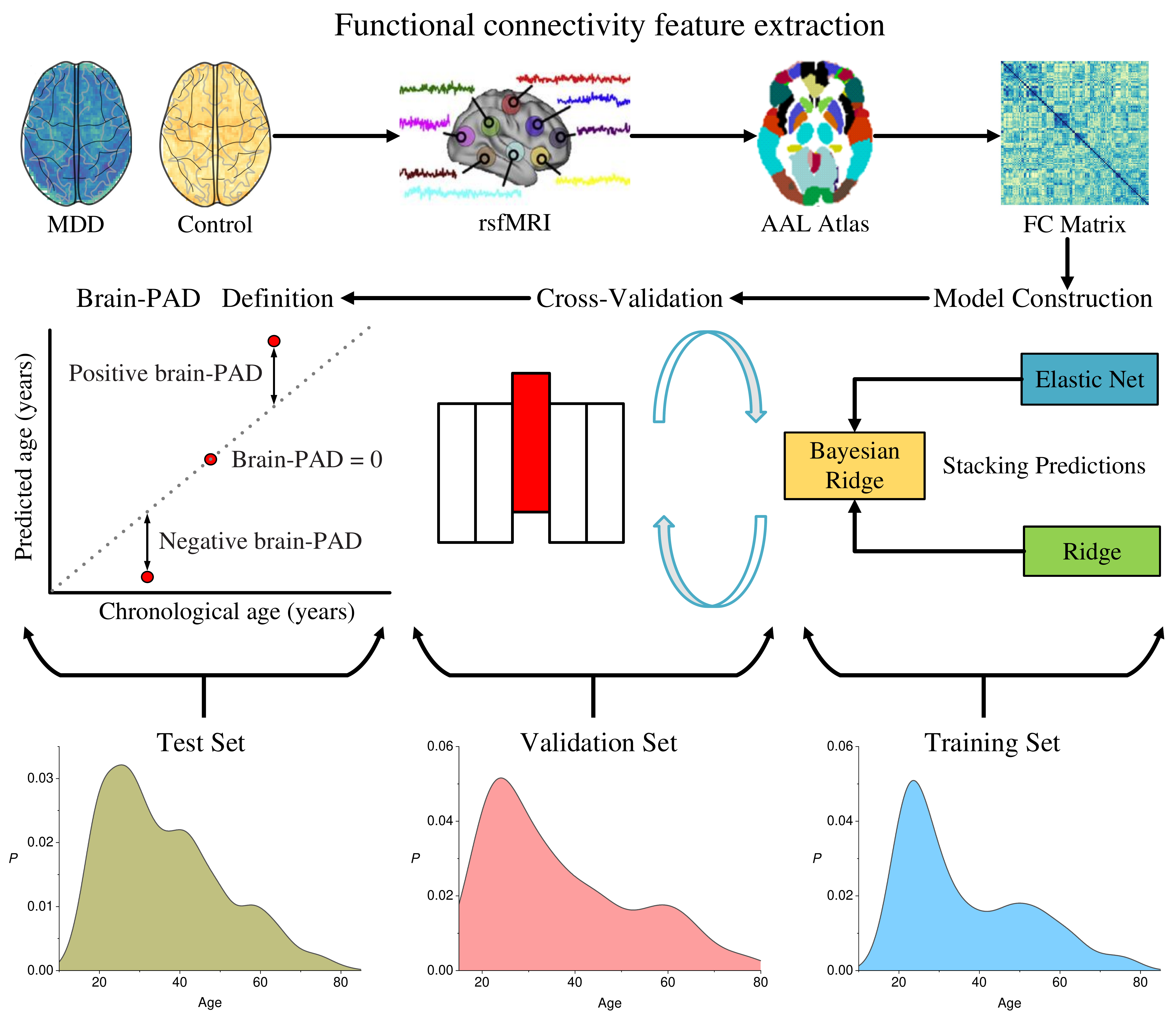} 
    \caption{The flow chart of the analysis. Time series for each subject are extracted from rsfMRI according to the AAL atlas. The Pearson correlation is used to get the functional connectivity matrix. Next, the upper triangle of the functional connectivity matrix is stretched into a one-dimensional vector to get each subject's feature vector. The feature vectors are concatenated to obtain input data matrices. A brain age prediction model is trained and five-fold cross-validated on the training set. Finally, brain-PAD scores are obtained by applying the model in the hold-out validation set and test set.}
    \label{flow chart}
\end{figure}

\subsection{Model training and evaluation}
To obtain the input feature for the model, we reshape the upper triangle of the whole-brain correlation matrix into a one-dimensional vector with 6670 elements. To determine the brain aging pattern in healthy individuals, we first train a brain age prediction model on the training set containing 1101 normal controls. Next, we utilize the model to estimate the brain age of 1276 MDD patients on the test set. The brain age prediction is first carried out by three classical supervised learning algorithms: elastic net \cite{ball2021individual,khundrakpam2015prediction}, bayesian ridge \cite{xu2019human}, and ridge regression \cite{han2020brain,niu2020improved,chung2018use}. Furthermore, we introduce a stacking model \cite{liem2017predicting} from ensemble learning \cite{couvy2020ensemble,da2020brain} to combine results from the three algorithms, which gives the best estimation results. The flow is shown in Fig. \ref{flow chart} and Supplementary Fig. \ref{stacking}. The four models all come to consistent conclusions in subsequent experiments. To avoid switching between different methods and make the flow of the paper more concise, we use the results from the stacking method in the main text. Analyses based on the other three algorithms are included in Supplementary Table \ref{Comparison of brain-PAD in different models}.

We evaluate our model performance in the control and MDD groups separately. We first evaluate the model on the entire training set with five-fold cross-validation. Then, the same model in each fold is used to predict the brain age of MDD patients on the entire test set. The performance of the four models is evaluated based on the following three metrics: mean absolute error (MAE), mean squared error (MSE) and mean coefficient of determination (R$^2$). All models are implemented through the Python-based sklearn package with all parameters set as the default value.

\subsection{Statistical analyses}
To determine whether brain aging is accelerated in MDD patients relative to controls, we split the entire controls to get a fixed training set and a validation set using the hold-out method \cite{levman2021hold}. While modest in size, this hold-out validation set consists of normal controls from all sites with the entire age span, providing an unbiased age representation of Rest-meta-MDD. We obtain a model on the training set and use it to estimate the brain age of the normal controls in the hold-out validation set and all MDD patients in the test set. The chronological age is subtracted from the estimated age to get the brain-PAD as the outcome variable for statistical analysis. The five-fold cross-validation is used to compare the overall performance of different models. The hold-out validation set is used as a normal control group for brain-PAD comparison. Due to factors such as regression dilution and non-Gaussian age distribution \cite{smith2019estimation}, we need to perform an age-bias correction. We apply a posthoc correction for the residual age effect on the test set \cite{niu2020improved,le2018nonlinear,liang2019investigating,cole2020multimodality,aycheh2018biological}. Following Peng et al. \cite{peng2021accurate}, we first train a linear regression model on the hold-out validation set to fit the predicted age with the chronological age labels. The slope and intercept of the fitting line are then used to correct the predicted age of the MDD patients in the test set. The systematic bias and the corresponding correction method are illustrated in Supplementary Fig. \ref{Comparison of predicted age before and after correction} and Fig. \ref{Correlation between bran-PAD and chronological age before and after correction}. We apply the univariate generalized linear model (GLM) with gender, diagnosis, age, and age$^2$ as covariates to explore the relationship between brain-PAD and clinical characteristics \cite{dunlop2021accelerated}. Furthermore, the two-sample t-test is used to compare the brain-PAD in different subgroups. Multiple comparisons are corrected by false discovery rates correction. The threshold for statistical significance is set at $\text{$p$} < 0.05$.

\section{Results}
\subsection{Model performance}
\begin{figure}[htb!]
    \centering
    \includegraphics[width=1\textwidth]{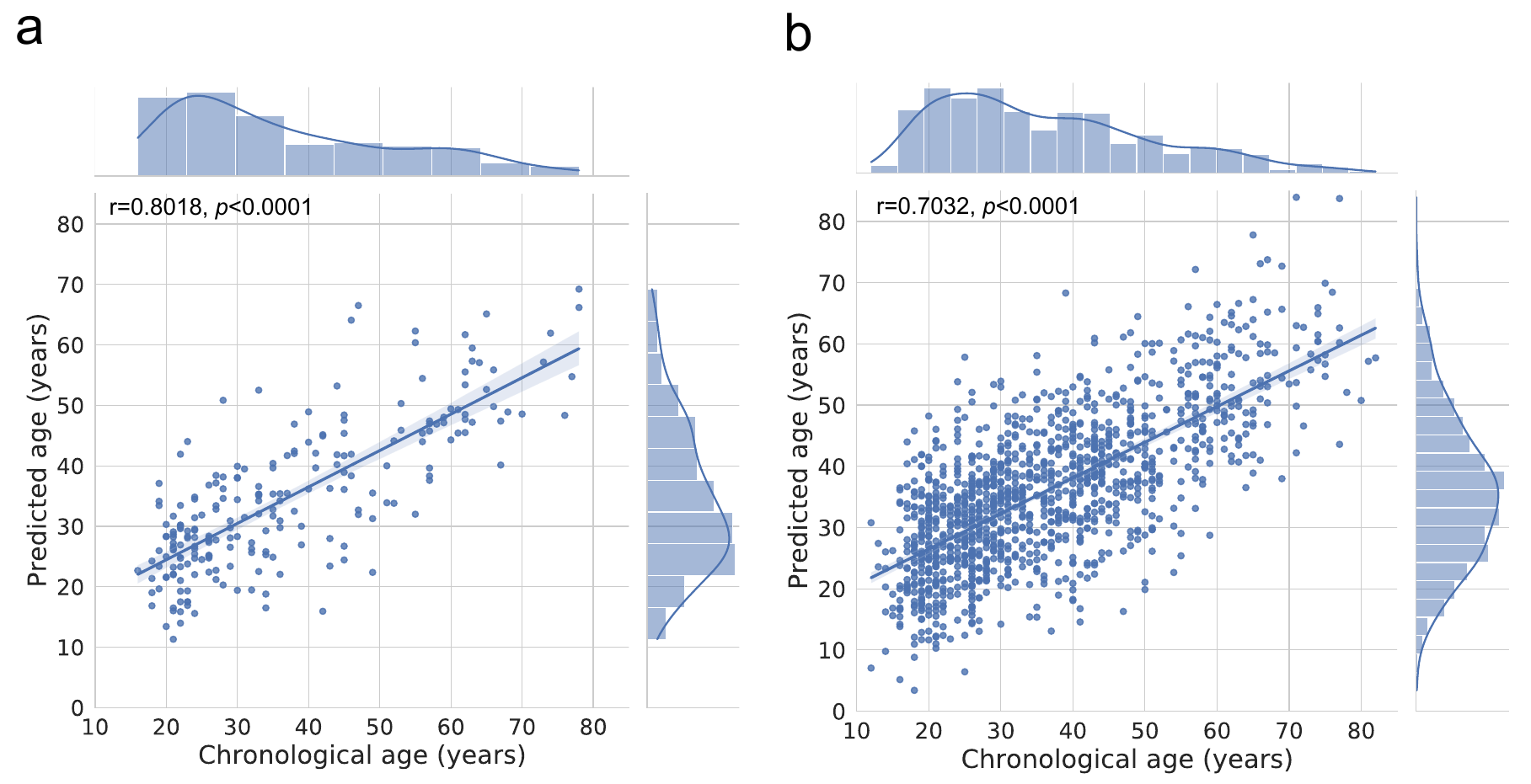}
    \caption{Correlation between the chronological age (x-axis) and predicted age (y-axis) obtained by the stacking model on the validation set (a) and test set (b). Statistical values are obtained from Pearson correlations two-sided test. The lines represent the regression results and the shades correspond to the 95$\%$ confidence interval.}
    \label{Correction}
\end{figure}

The models obtained from each fold of the training set are used to estimate the brain age of individuals for the rest of the controls in the validation set as well as the MDD patients in the test set. Table \ref{tab:Performance of different models in Rest-meta-MDD data} shows the performance of four models with 882 training subjects, 219 validation subjects, and 1276 test subjects. Among the three classical machine learning algorithms, the bayesian ridge achieves the best performance. But the stacking model with ensemble learning outperforms all of them, giving rise to the lowest MAE and MSE in both the validation and test set. More comparative results can be found in Supplementary Table \ref{Performance of different models}. The correlation between chronological age and predicted age is presented in Fig. \ref{Correction}.

\begin{table}[htb!]
\centering
\caption{Performance of four models.}
\label{tab:Performance of different models in Rest-meta-MDD data}
\begin{tabular}{cccc}
\toprule
\hline\multicolumn{4}{c}{Validation set}                                      \\ \midrule\hline
Model          & MAE             & MSE                & R$^2$              \\
Elastic Net    & 8.2327 ± 0.4608 & 103.7454 ± 10.9086 & 0.5670 ± 0.0486 \\
Ridge          & 8.7749 ± 0.5662 & 127.4526 ± 17.7452 & 0.4691 ± 0.0665 \\
Bayesian Ridge & 7.8057 ± 0.4420 & 97.4546 ± 10.5659  & 0.5934 ± 0.0440 \\
Stacking       & 7.7287 ± 0.5547 & 95.3625 ± 11.8727  & 0.6026 ± 0.0456 \\ \midrule
\hline\multicolumn{4}{c}{Test set}                                            \\ \midrule\hline
Elastic Net    & 8.4156 ± 0.0582 & 110.2473 ± 1.2672  & 0.4839 ± 0.0059 \\
Ridge          & 9.4921 ± 0.2447 & 143.8590 ± 6.8188  & 0.3265 ± 0.0319 \\
Bayesian Ridge & 8.3817 ± 0.0609 & 110.6582 ± 1.2202  & 0.4820 ± 0.0057 \\
Stacking       & 8.3055 ± 0.0535 & 108.4852 ± 1.2633  & 0.4837 ± 0.0071 \\ \bottomrule\hline
\end{tabular}
\end{table}

\begin{figure}
    \centering
    \includegraphics[width=1\textwidth]{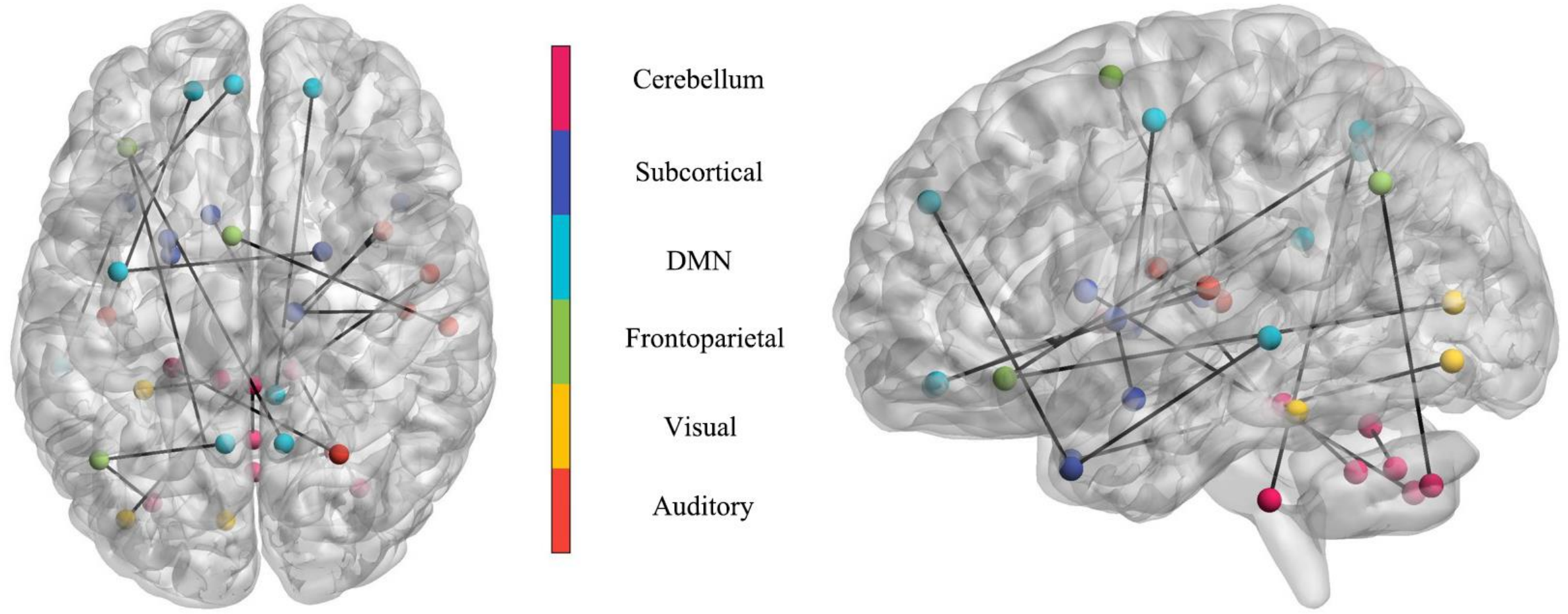}
    \caption{The top 20 most important functional connectivity features obtained from the bayesian ridge model. These features consist of functional connectivity between the cerebellum superior and vermis8, medial superior frontal gyrus and middle temporal gyrus, amygdala and lenticular nucleus putamen, orbital inferior frontal gyrus and precuneus, angular gyrus and precuneus.}
    \label{Feature Importance}
\end{figure}

\subsection{The relative feature importance for normal controls}
\begin{figure}[htb!]
\renewcommand{\thefigure}{S6}
\centering
\includegraphics[width=0.55\textwidth]{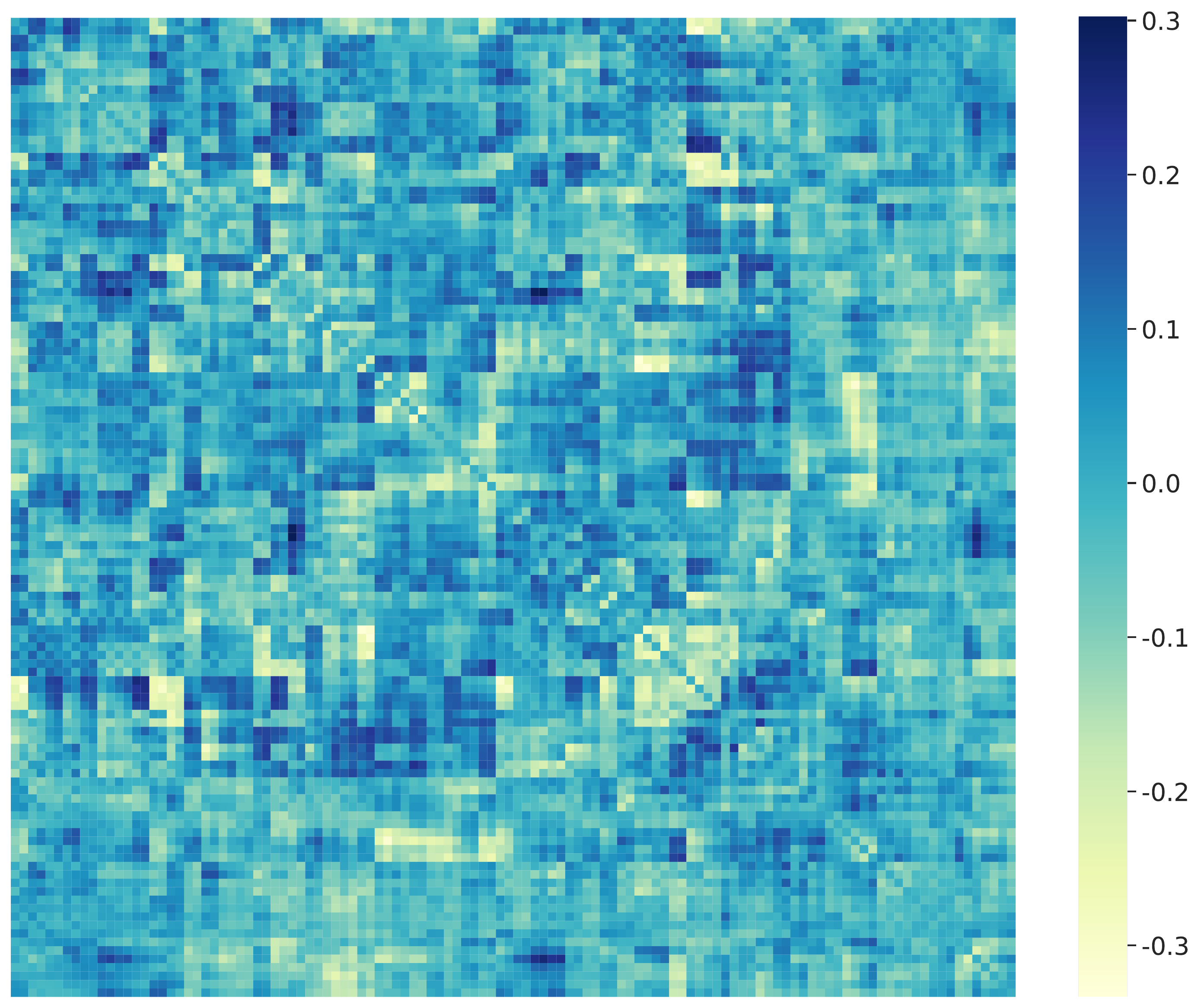}
\caption{Correlation between the functional connectivity features and the chronological age.  Among all the total 6670 functional connectivity features, 3196 features show positive correlations with age mean $\text{correlation} = 0.0645 \pm 0.0495$, $\text{range} (6.1691e-05, 0.3017)$. 3474 features show negative correlations with age mean $\text{correlation} = -0.0691 \pm 0.0545$ with $\text{range} (-0.3334, 3.7818e-05)$.}
\label{Correlation between the functional connectivity features and the chronological age}
\end{figure}

We calculate the correlation between the functional connectivity features and the chronological age (Supplementary Fig. \ref{Correlation between the functional connectivity features and the chronological age}). Among all the total 6670 functional connectivity features, 3196 features show positive correlations with age (mean $\text{correlation} = 0.0645 \pm 0.0495$, $\text{range} (6.1691e-05, 0.3017)$). 3474 features show negative correlations with age (mean $\text{correlation} = -0.0691 \pm 0.0545$, $\text{range} (-0.3334, 3.7818e-05)$). In particular, the most positive correlation is found for the precentral gyrus \cite{kovalev2003gender} - heschl gyrus \cite{amoroso2019deep}. The most negative correlation is found for the median cingulate and paracingulate gyry - inferior parietal gyrus, excluding supramarginal and angular gyri \cite{macdonald2008increased,rivera2005developmental}. In addition, we identify brain regions that the machine learning algorithm considers to be significant in the brain age estimation using the feature importance \cite{hellyer2013individual}. The feature importance values are normalized to give the top 20 functional connectivity features (Fig. \ref{Feature Importance}). The main brain regions include the cerebellum \cite{tiemeier2010cerebellum} superior and vermis8, the medial superior frontal gyrus and middle temporal gyrus \cite{tomoda2012pseudohypacusis}, the amygdala and lenticular nucleus putamen. These brain regions are associated with brain development and atrophy, which are consistent with previous studies. The detailed feature importance values are shown in Supplementary Fig. \ref{The top 20 feature importance value}.

\begin{figure}[htb!]
    \centering
    \includegraphics[width=1\textwidth]{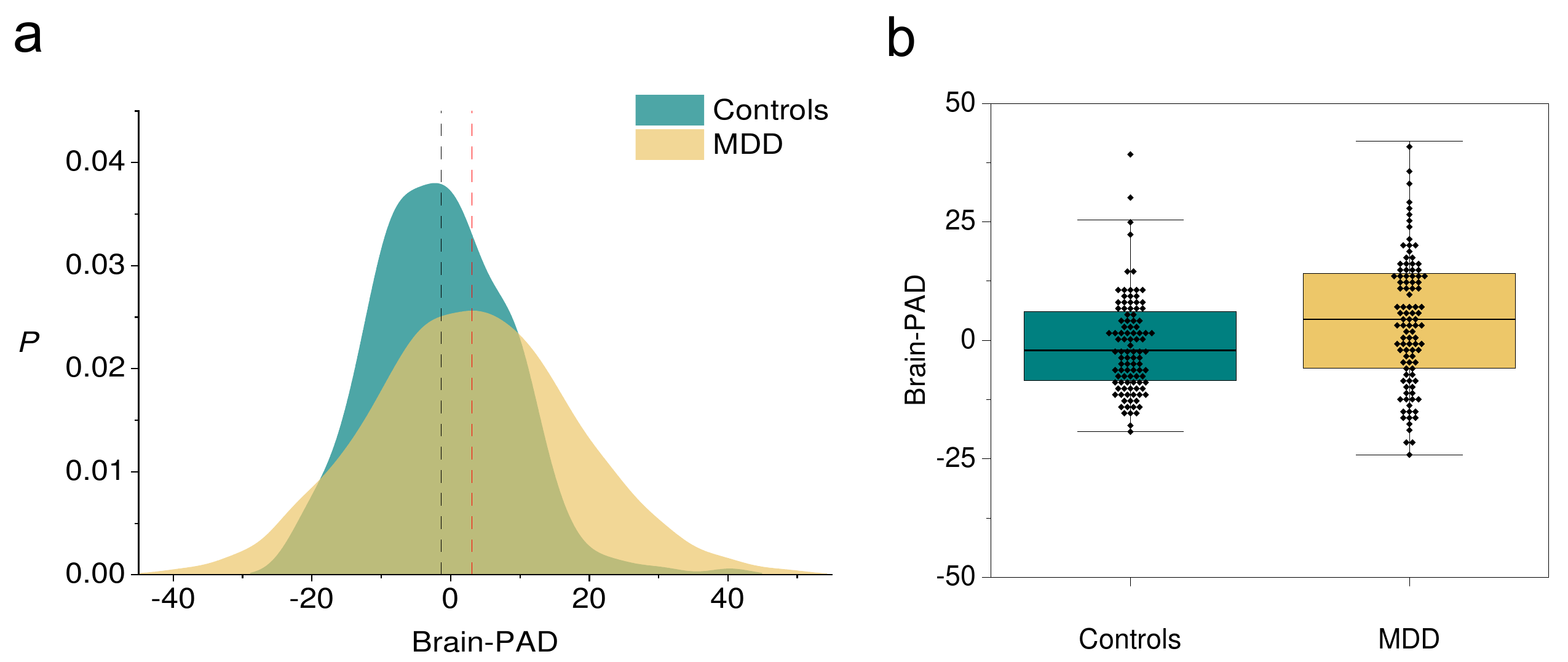}
    \caption{Brain predicted age difference in controls and MDD patients. Group analysis shows that the corrected brain-PAD is significantly higher in MDD patients than in controls ($\text{$p$} < 0.0001$, $\text{Cohen's $d$} = 0.35$, $\text{95\% CI}:1.86 - 3.91$).}
    \label{Brain-PAD}
\end{figure}

\subsection{Accelerated functional brain aging in MDD}
We compare the resulting brain-PAD scores of the MDD patients with the controls in the hold-out validation set to determine whether brain aging is accelerated in MDD patients. Overall, the brain-PAD score before age-bias correction is -1.3731 (SD 9.91) years in the control group and -0.0712 (SD 10.56) years in MDD patients. After applying an age-bias correction procedure, the brain-PAD is $+4.43$ years higher in MDD patients than in normal controls ($\text{$p$} < 0.0001$, $\text{Cohen's $d$} = 0.35$, $\text{95\% CI}:1.86 - 3.91$), which is shown in Fig. \ref{Brain-PAD}. Although different estimations are obtained through different models, results from the other three models all demonstrate a consistent pattern that MDD patients have a statistically significant higher brain-PAD scores compared to controls. In addition, GLM shows significant main effects for age ($\text{$p$} < 0.001$), age$^2$ ($\text{$p$} < 0.002$) and diagnosis ($\text{$p$} < 0.0001$), but not for gender (Table \ref{tab:main effects in gender, Diagnosis, Age and Age$^2$}).

\begin{table}[htb!]
\centering
\caption{Parameter estimates for all main effects and significant
interactions in gender, diagnosis, age and age$^2$.}
\label{tab:main effects in gender, Diagnosis, Age and Age$^2$}
\begin{tabular}{@{}lcccccr@{}}
\hline\toprule
          & Coef    & SE    & $z$      & $p$     & {[}0.025 & 0.975{]} \\ \midrule\hline
Intercept & 7.6400  & 2.570 & 2.972  & 0.003 & 2.602    & 12.678   \\
Gender    & -0.6052 & 0.771 & -0.785 & 0.432 & -2.116   & 0.905    \\
Diagnosis & 4.7237  & 1.045 & 4.519  & 0.000 & 2.675    & 6.772    \\
Age       & -0.4421 & 0.129 & -3.433 & 0.001 & -0.695   & -0.190   \\
Age$^2$     & 0.0046  & 0.002 & 3.048  & 0.002 & 0.002    & 0.008    \\ \bottomrule\hline
\end{tabular}
\end{table}

\begin{table}[htb!]
\centering
\caption{Parameter estimates for all main effects and significant
interactions in other clinical characteristics.}
\label{tab:main effects in other clinical characteristics}
\begin{tabular}{@{}lcccccr@{}}
\hline\toprule
           & Coef    & SE    & $z$      & $p$     & {[}0.025 & 0.975{]} \\ \midrule\hline
Intercept  & 1.8688  & 5.087 & 0.367  & 0.713 & -8.102   & 11.839   \\
Gender     & -1.2863 & 1.255 & -1.025 & 0.305 & -3.746   & 1.173    \\
Eposide    & 2.2518  & 1.532 & 1.470  & 0.141 & -0.750   & 5.254    \\
Medication & 2.9454  & 1.291 & 2.282  & 0.023 & 0.415    & 5.475    \\
Age        & -0.0285 & 0.252 & -0.113 & 0.910 & -0.522   & 0.465    \\
Age$^2$        & -0.0007 & 0.003 & -0.219 & 0.827 & -0.007   & 0.006    \\
Education  & -0.0022 & 0.156 & -0.014 & 0.989 & -0.307   & 0.303    \\
Month      & 0.0060  & 0.012 & 0.511  & 0.609 & -0.017   & 0.029    \\ \bottomrule\hline
\end{tabular}
\end{table}

\begin{figure}[htb!]
    \centering
    \includegraphics[width=1\textwidth]{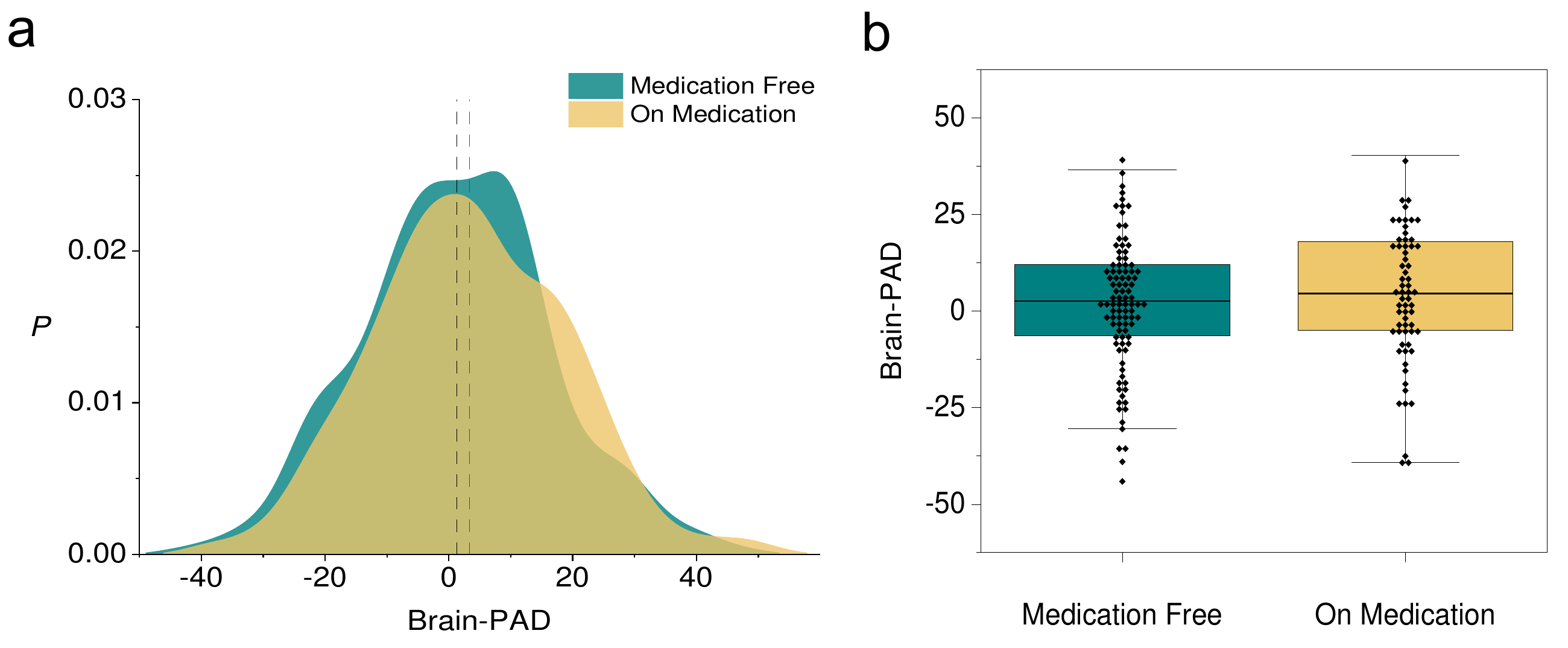}
    \caption{Brain Predicted age difference in antidepressant users and medication-free patients. Group analysis shows that brain-PAD is higher in MDD patients who are taking medication ($+3.38$ years, $\text{95\% CI}:0.91-4.65$) than in those medication-free ones ($+1.29$ years, $\text{95\% CI}:-0.41-3.10$). The difference passes the statistic significant threshold ($\text{$p$} = 0.0499$).}
    \label{Medication}
\end{figure}

\subsection{Brain-PAD comparison for clinical characteristics}
\begin{figure}[htb!]
\renewcommand{\thefigure}{S8}
\centering
\includegraphics[width=0.8\textwidth]{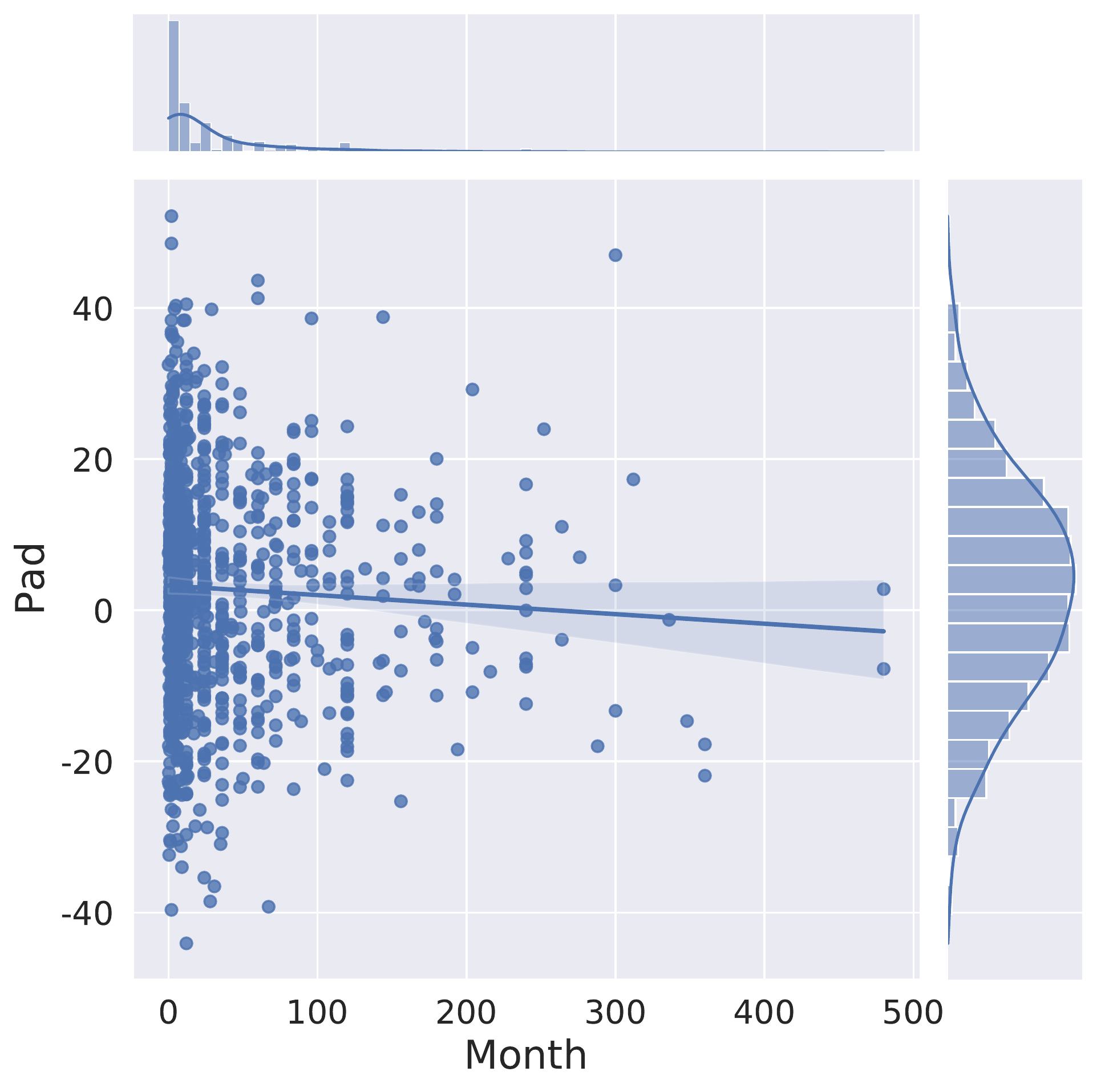}
\caption{The correlation between brain-PAD scores and illness duration months in patients with depression. There is a negative correlation between the brain-PAD and illness duration ($\text{Spearman $R$} = -0.067$, $\text{$p$} = 0.03$).}
\label{The correlation between brain-PAD scores and illness duration in patients with depression}
\end{figure}

To explore the association between brain-PAD scores and clinical characteristics, we use the GLM to fit the brain-PAD of MDD patients with the following explanatory variables: sex, medication status, episode status, education years, and illness duration months (Table \ref{tab:main effects in other clinical characteristics}). The medication status ($\text{$p$} = 0.023$) has a main effect on the brain-PAD scores of MDD patients. We further apply a two-sample t-test to determine whether the brain-PAD mean value in antidepressant users and medication-free patients are significantly different from each other (Fig. \ref{Medication}). Brain-PAD is $+2.09$ years higher ($\text{$p$} = 0.0499$, $\text{Cohen's $d$} = 0.13$) in antidepressant users than in medication-free ones. Comparisons of other subgroups (sex, episode status) with controls can be found in Supplementary Table \ref{Comparison of brain-PAD between different clinical characteristics}. While significant differences are observed in all MDD subgroups compared to normal controls, posthoc comparisons of brain-PAD in other clinical characteristics do not demonstrate any significant differences between MDD subgroups except for the medication status. For the two continuous-type clinical characteristics (education years and illness months), we divide the subgroups according to their medians (both are 12 in this study) for brain-PAD comparisons (Supplementary Table \ref{Comparison of brain-PAD in education years and illness duration months}). Overall, MDD patients with fewer than 12 years of education have a 2.28 years higher brain-PAD than those with greater than or equal to 12 years of education ($\text{$p$} = 0.00679$). Brain-PAD of MDD patients with fewer than 12 months of illness is 1.69 years higher than that in patients with greater than or equal to 12 months of illness. We also calculate correlations between brain-PAD scores and illness months, education years, and HDRS scores separately. Only illness duration is found to be significantly correlated with brain-PAD scores ($\text{Spearman $R$} = -0.067$, $\text{$p$} < 0.05$, Supplementary Fig. \ref{The correlation between brain-PAD scores and illness duration in patients with depression}).

\section{Discussion}
Biological aging can be defined as a progressive process of decline involving multiple organ systems. While all individuals age chronologically at the same rate, the rate of their biological aging varies from one to the other \cite{elliott2021disparities}. Resting-state functional MRI is developed as a common approach to interrogate the myriad of functional systems in the brain without the constraints of any prior assumptions \cite{biswal2010toward}. Machine learning algorithms based on functional connectivity and the availability of large-scale reliable samples allow us to develop generalized models to estimate the brain age of individual subjects \cite{franke2019ten}. Here, we make use of the Rest-Meta-MDD consortium from China to verify the accelerated brain aging in MDD patients, which is previously observed in Caucasian participants using structural MRI information. We apply four machine learning algorithms based on functional connectivity features to estimate the brain age of individuals with the entire adult lifespan (12-82 years). We observe manifestly accelerated brain aging in 1276 MDD patients. Furthermore, we compare brain-PAD scores between MDD subgroups divided according to clinical characteristics such as medication status and episode status. We confirm that the conclusion drawn in this paper is not algorithm sensitive as results from different algorithms lead to the same conclusion.

Our study benefits from a reliable experimental design. The dataset contains 24 cohorts so the potential site effect is effectively avoided. Instead of using samples from some independent sites as the fixed validation set, we randomly select samples from all sites to constitute the training set and validation set. In this way, the generalizability of the model is improved and the model outcomes are evaluated more objectively \cite{orru2020machine,cai2020generalizability}. Moreover, we split the normal controls into a fixed training set and a hold-out validation set. We compare the brain-PAD scores of the controls in this hold-out validation set to the MDD patients in the test set. As the validation set is not involved in the development of the brain age prediction model, the risk of overfitting is effectively prevented \cite{shim2021inflated}. The application of four different machine learning algorithms allows us to further validate the consistency of the patterns observed.

\begin{figure}[htb!]
\renewcommand{\thefigure}{S5}
\centering
\includegraphics[width=0.75\textwidth]{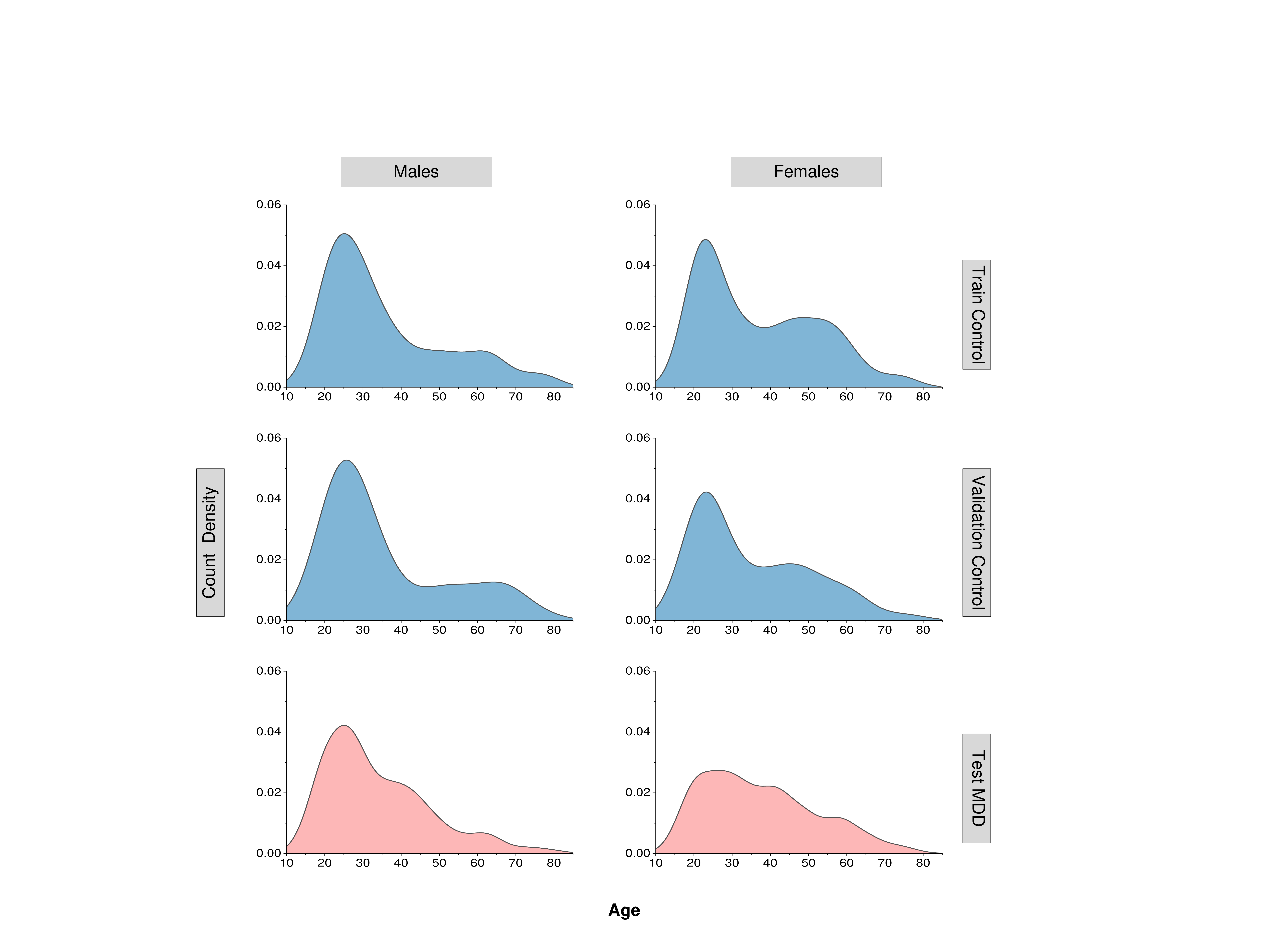}
\caption{Age distribution in Rest-meta-MDD. The Rest-meta-MDD contains 2377 participants (12-82 years old).}
\label{Age distribution in Rest-meta-MDD}
\end{figure}

Although multiple studies are carried out on relatively small samples, conclusions drawn from larger samples tend to be more reliable. First, machine learning algorithms are sensitive to sample size \cite{arbabshirani2017single,gianfrancesco2018potential,varoquaux2018cross,vabalas2019machine}. The small size of samples brings a bigger prediction error and a higher risk of overfitting \cite{marek2022reproducible}. Moreover, larger samples tend to contain subjects with a wider age distribution. While a wide range of ages makes the estimation challenging, it effectively increases the generalizability of the conclusion \cite{cole2019quantification}. In this study, we make use of the Rest-meta-MDD consortium, which is the largest rsfMRI database of MDD patients. To the best of our knowledge, only one study from ENIGMA uses more subjects than ours, which contains a total of 6989 subjects aged from 18 to 75 years old \cite{han2020brain,schmaal2020enigma}. But compared with ENIGMA, subjects in Rest-meta-MDD have a bigger life span, ranging from 12 to 82 years old (see the age distribution of the two datasets in Supplementary Fig. \ref{Age distribution in ENIGMA} and Fig. \ref{Age distribution in Rest-meta-MDD}). The sufficiently large samples with a wider age span lead to similar conclusions drawn in ENIGMA.

Our results show a $+4.43$ years gap in terms of the brain-PAD between MDD patients and normal controls (with a Cohen’s $d$ effect sizes size of 0.35) at the group level. Compared to previous studies on accelerated brain aging in MDD patients, such as the one by Koutsouleris et al. \cite{koutsouleris2014accelerated} ($+4.0$ years, $\text{N} = 104$), Han et al. \cite{han2020brain} ($+1.16$ years, $\text{N} = 2675$), Dunlop et al. \cite{dunlop2021accelerated} ($+2.11$ years, $\text{N} = 112$), Han et al. \cite{han2021stage} ($+0.586$ years, $\text{N} = 195$), our results demonstrate a higher brain-PAD. We speculate that this may be related to the stigmatization of depression in traditional Chinese culture \cite{georg2008stigma}. Some studies suggest that compared to Caucasians, Chinese tend to deny the existence of depression \cite{parker2001depression}. Consequently, the level of depression tends to be significantly elevated when MDD is diagnosed \cite{young2010depression,ai2015ethnic,krieg2018comparing}.

We find the significant main effects of age, age$^2$, and diagnosis in our regression analysis. In particular, the main effect of medication status on brian-PAD is observed after the inclusion of other clinical characteristics, which is in line with the finding by Sacchet et al. \cite{sacchet2017accelerated}. The comparisons between subgroups of MDD patients show that antidepressant users have a statistically significant higher brain-PAD than medication-free patients ($+2.09$ years, $\text{$p$} = 0.0499$, $\text{Cohen's $d$} = 0.134483$). Explanations for this phenomenon are discussed by Han \cite{han2020brain}. The antidepressant users are likely to have a more severe or chronic course of the disorder at the time of scanning. Therefore, the larger brain-PAD scores in antidepressant users may be confounded by clinical standards recommending antidepressant use mainly for severe or chronic MDD \cite{schmaal2017cortical}. In other words, patients with milder symptoms tend not to take antidepressants \cite{schmaal2016subcortical}. To fully understand the adaptation of brain-PAD in response to pharmacotherapy, randomized controlled intervention studies are needed which require more information on the clinical use of antidepressants, such as the dosage and duration. It is also noteworthy that the $p$-value in the study by Han \cite{han2020brain} is slightly above 0.05 whereas in our study it is slightly below the threshold of statistical significance. But it is still near the boundary of the threshold line. We honestly report the result, and we also admit that it is far early to draw any conclusion based on this statistic.

Similarly, consistent with the finding obtained by Han et al. \cite{han2021stage} using structural brain MRI, we observe a higher brain-PAD in first-episode patients ($+4.19$ years, $\text{N} = 538$) than in recurrent patients ($+2.56$ years, $\text{N} = 282$). We believe the same explanation can be applied. First, as pointed out in previous studies and observed in our work, there is a negative correlation between the brain-PAD and illness duration ($\text{Spearman $R$} = -0.067$, $\text{$p$} < 0.05$). Furthermore, recurrent patients have a longer illness duration than first-episode patients. The median illness duration in recurrent patients (60 months) is 10 times greater than in first-episode patients (6 months) in our data. The corresponding median of brain-PAD score in recurrent patients (0.58 years) is 3.08 years smaller than in first-episode patients (3.66 years). The combination of the two effects gives rise to a higher brain-PAD in first-episode patients than that in recurrent patients. It is implied that there may be a clinically unstable period in first-episode patients. As more treatment is given, patients may become more stable in brain functioning. Hence the brain-PAD decreases with the illness duration \cite{han2021stage}. But such a hypothesis needs more clinical information to be further verified through longitudinal studies.

Our results extend the generalizability of accelerated brain aging in MDD patients using the rsfMRI feature of Chinese participants. But several limitations should be considered. Although a standardized preprocessing pipeline is employed at all sites before the aggregation group analysis, some subjects still show measurement bias and missing values in the scan. We address this problem by applying various standardization methods to the features. Although the prediction error is within the control, these operations may still bring impacts on the final results. Next, multiple brain atlas could be considered to obtain the functional connectivity features. Different functional connectivity will have an impact on the subsequent analysis. Furthermore, different features and models could also have a dramatic effect on the final results. Several studies report the great potential of the multimodal features \cite{cherubini2016importance,de2020multimodal,rokicki2021multimodal} and deep learning algorithms \cite{jonsson2019brain,schulz2020different,abrol2021deep} in neuroimaging research. More comparisons of neuroimaging features and models are needed in the future to produce more convincing conclusions. Besides, all participants in Rest-meta-MDD are Chinese, the generalizability of our model to other ethnic/racial and cultural backgrounds remained to be explored. Finally, aging is a continuous process, yet few current studies address longitudinal investigations of brain aging, including stage-by-stage analyses of MDD to explore trends in brain-PAD with age to understand the progressive effects of the aging process. More clinical features are still desired in the future to determine the clinical significance of measuring brain-PAD and whether it can be considered as a clinically essential biomarker.

\section*{Acknowledgments}
This work is supported by Industry-University-Research Innovation Fund for Chinese Universities (No. 2021ALA03016), University Innovation Research Group of Chongqing (No. CXQT21005). We thank the REST-meta-MDD consortium for sharing the data.

\clearpage

\bibliographystyle{naturemag}
\bibliography{mybibfile}

\section*{Supplementary Information}

\subsection*{S1. Data acquisition}
The REST-meta-MDD consortium contains a total of 25 neural cohorts from incoming groups in China. Site 4 (including 24 MDD patients and 24 controls) is excluded in that it is a duplicate of site 14 (detected during the preparation of REST-meta-MDD data for open sharing). We construct a functional connectivity network for each subject based on the AAL atlas using time series data from subjects at the remaining 24 sites. The specific information for each site is shown in Table \ref{Samples of selected sites} and Tabel \ref{Data acquisition parameters of selected sites}. We take the step of adjusting the characteristics of all subjects by various normalization methods. Although these operations may have a slight effect on the final results, we consider these errors to be within reasonable control.

\subsection*{S2. Data preprocessing and quality control}

The initial 10 functional images are discarded with the slice acquisition timing discrepancies and the head motion is performed. Linear trends, friston 24 head motion parameters, the white matter signal, and the cerebrospinal fluid signal are regressed out from the functional signal as nuisance covariates. Our quality control is more relaxed compared to previous studies. We only remove duplicate samples from site 4 and three samples with abnormal age, resulting in 1276 MDD patients and 1101 controls. Various types of subjects are removed in Yan et al. \cite{yan2019reduced} and Yang et al. \cite{yang2021disrupted}, such as subjects from small sample sites, subjects younger than 18 or older than 65, subjects with vacancies in clinical characteristics, etc. We believe that the age distribution of the training set needs to be made as wide as possible for the model to have better generalizability. At the same time, the proposed algorithm predicts the brain age of subjects based on rsFC features, and the vacancies in clinical features do not affect the algorithm to predict the brain age of subjects based on their rsFC features.

\subsection*{S3. Stacking model}
\begin{figure}[htb!]
\renewcommand{\thefigure}{S1}
\centering
\includegraphics[width=1\textwidth]{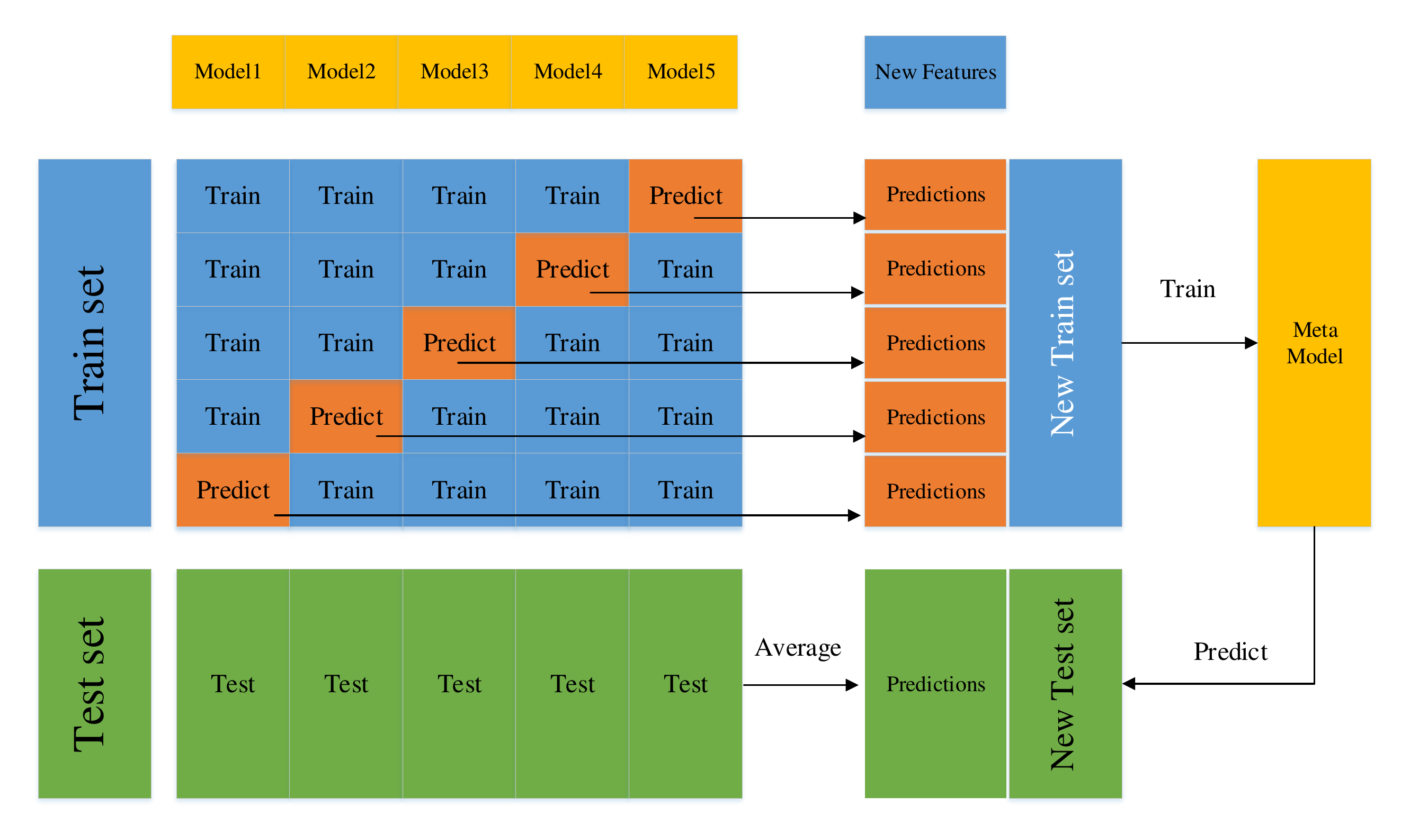}
\caption{Schematic diagram of two levels stacking.}
\label{stacking}
\end{figure}

Stacking model is an ensemble learning approach that combines different prediction models in a single model, working at levels or layers. This approach aims to minimize the errors of generalization by reducing the bias of its generalizers. Considering a stacking approach using two levels (level-0 and level-1) in Fig. \ref{stacking}. In level-0, diverse base models are trained, and the prediction of the response variable for each one is performed subsequently. These forecasts are used as a new input feature for the level-1 model, which is also called meta-model \cite{shamaei2016suspended}. At the same time, the base classifiers need to predict the test set, and the predictions are averaged as a new test set for the meta classifier to predict the desired result. 

\clearpage

\subsection*{S4. Correcting the age dependence of brain-PAD}
\begin{figure}[htb!]
\renewcommand{\thefigure}{S2}
\centering
\includegraphics[width=0.8\textwidth]{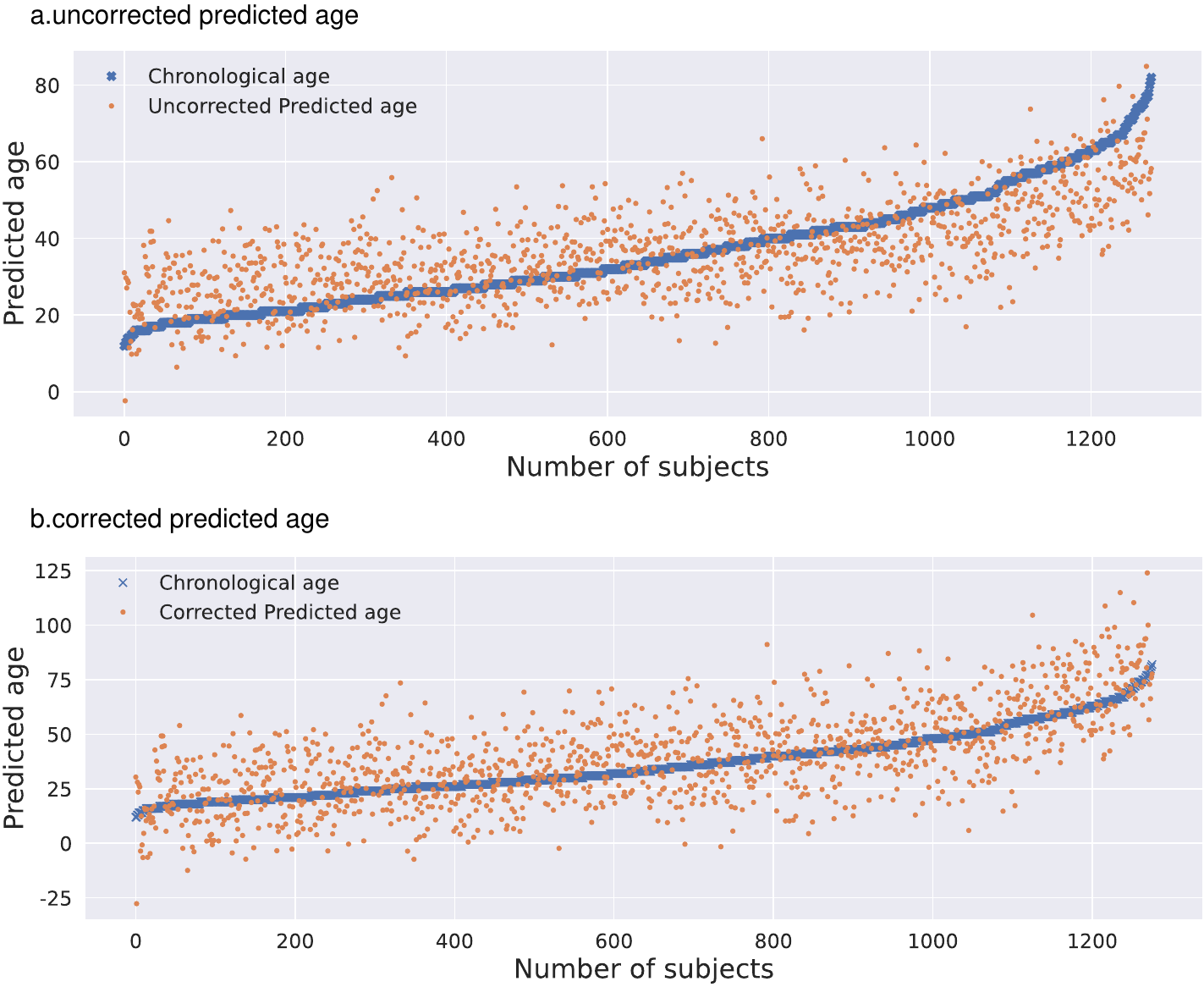}
\caption{Comparison of predicted age before and after correction.}
\label{Comparison of predicted age before and after correction}
\end{figure}

\begin{figure}[htb!]
\renewcommand{\thefigure}{S3}
\centering
\includegraphics[width=0.9\textwidth]{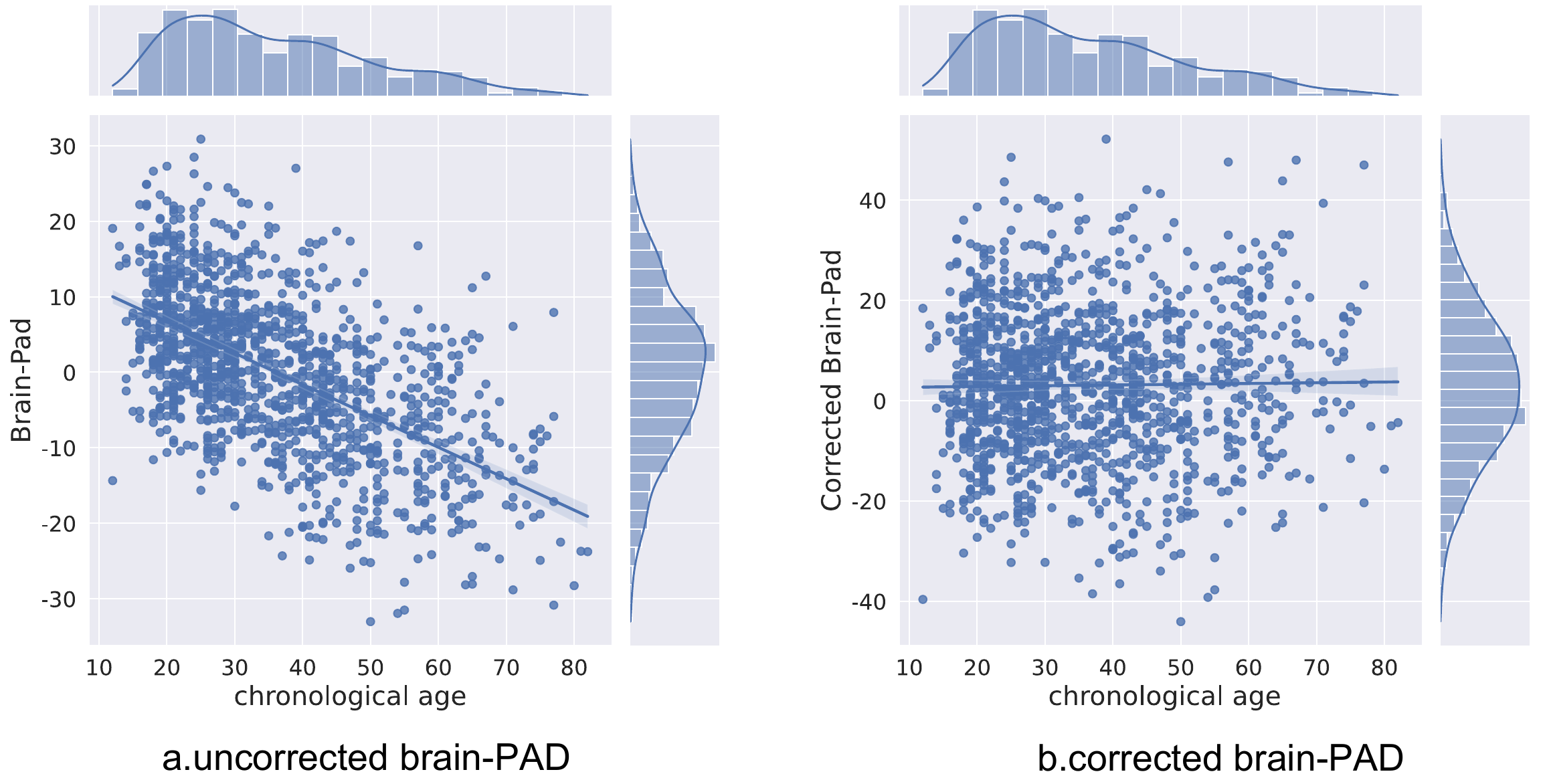}
\caption{Correlation between bran-PAD and chronological age before and after correction.}
\label{Correlation between bran-PAD and chronological age before and after correction}
\end{figure}

As shown in Fig. \ref{Comparison of predicted age before and after correction}a and Fig. \ref{Correlation between bran-PAD and chronological age before and after correction}a, following many existing studies, we observe a significant correlation between brain-PAD and chronological age. It means that the predicted age tends to be biased towards the mean age of the cohort, implying that younger subjects will be predicted to be older and vice versa. To correct for the presence of regression dilution bias in age prediction, we train a linear regression model to fit the predicted age on the validation set with age labels set aside. Then the slope and intercept of this model are used to correct for the predicted age from the test set. The corrected predicted age and the correlation between corrected brain-PAD and chronological age are shown in  Fig. \ref{Comparison of predicted age before and after correction}b and Fig. \ref{Correlation between bran-PAD and chronological age before and after correction}b. Following Peng et al. \cite{peng2021accurate}, we define $y$ to be chronological age and $x$ to be the predicted age, and we fit a linear regression (Equation \ref{Equation 1}) to the hold-out validation set (with labels). The corrected predicted age is estimated by Equation \ref{Equation 2}. 

\begin{equation}
x = ay + b
\label{Equation 1}
\end{equation}

\begin{equation}
\hat{x} = \frac{x-b}{a}
\label{Equation 2}
\end{equation}

This method requires (at the point of estimating $a$ and $b$ from $x$ and $y$) that the chronological ages are known. For the label-missing (final evaluation) test set, we assume that $a$ and $b$ are generalizable, and use the coefficients previously fit in the hold-out validation set to estimate the brain age.

\clearpage

\begin{figure}[htb!]
\renewcommand{\thefigure}{S4}
\centering
\includegraphics[width=0.65\textwidth]{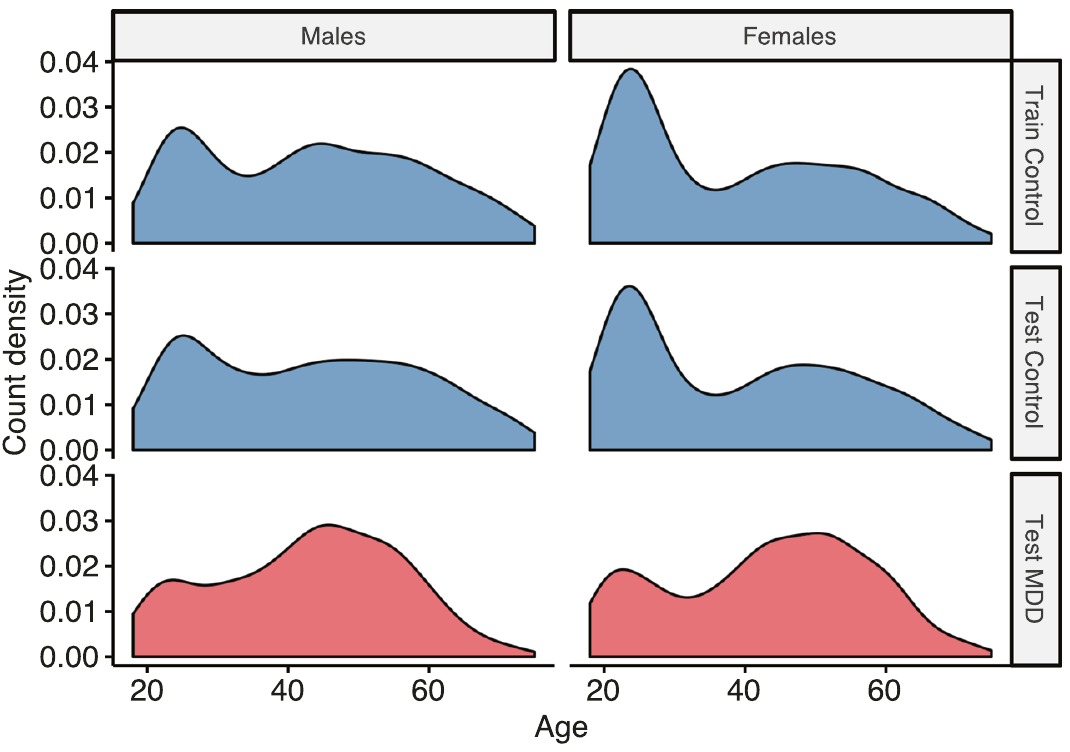}
\caption{Age distribution in ENIGMA. The ENIGMA MDD Working Group contains 6989 participants (18-75 years old).}
\label{Age distribution in ENIGMA}
\end{figure}

\begin{figure}[htb!]
\renewcommand{\thefigure}{S7}
\centering
\includegraphics[width=0.7\textwidth]{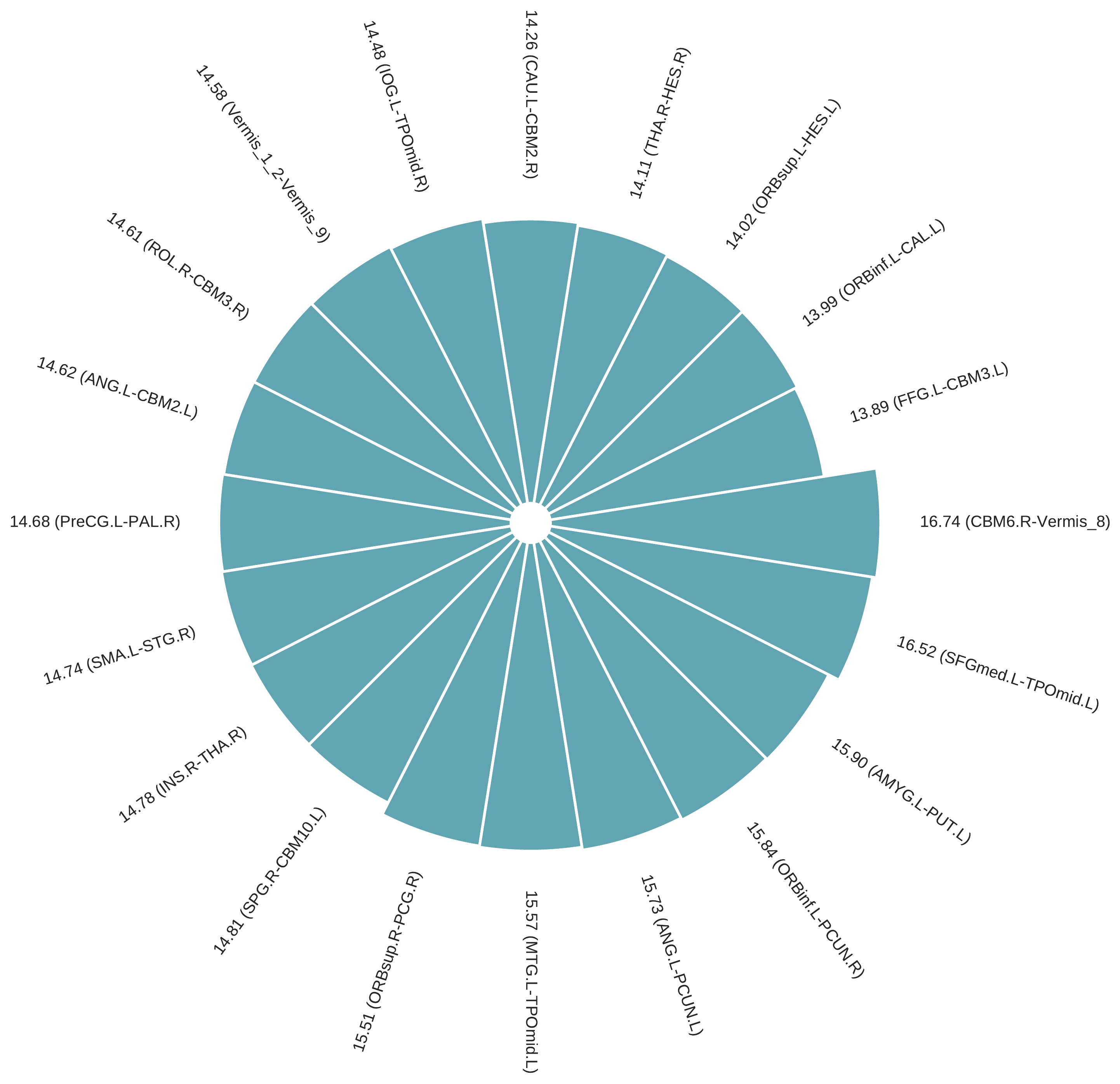}
\caption{The top 20 feature importance value. The values of all feature importance are normalized by a factor of 100. The maximum value is 16.736843 for the correlation between cerebellum superiorR and vermis8.}
\label{The top 20 feature importance value}
\end{figure}

\begin{sidewaystable}[hbt!]
\renewcommand{\thetable}{S1}
\centering
\caption{Samples of selected sites.}
\label{Samples of selected sites}
\begin{tabular}{@{}llll@{}}
\hline
\toprule
Serial number & Research groups                                                                                                                                                           & MDD patients (n) & NCs (n) \\ \midrule \hline
1              & \begin{tabular}[c]{@{}l@{}}National Clinical Research Center for Mental Disorders, Peking University\end{tabular} & 74               & 74      \\
2              & \begin{tabular}[c]{@{}l@{}}The Affiliated Guangji Hospital of Soochow University\end{tabular}            & 30               & 30      \\
3              & The Second Xiangya Hospital of Central South University                                                                                                                   & 27               & 37      \\
5              & Department of Psychiatry,Shanghai Jiao Tong University School of Medicin                                                                                                  & 13               & 11      \\
6              & Department of Psychiatry,Shanghai Jiao Tong University School of Medicine                                                                                                 & 15               & 15      \\
7              & Sir Run Run Shaw Hospital,Zhejiang University School of Medicine                                                                                                          & 38               & 49      \\
8              & Department of Psychiatry,First Affiliated Hospital,China Medical University                                                                                               & 75               & 75      \\
9              & The First Affiliated Hospital of Jinan University                                                                                                                         & 50               & 50      \\
10             & First Hospital of Shanxi Medical University                                                                                                                               & 50               & 33      \\
11             & Department of Psychiatry,The First Affiliated Hospital of Chongqing Medical University                                                                                    & 32               & 29      \\
12             & Department of Psychiatry,The First Affiliated Hospital of Chongqing Medical University                                                                                    & 32               & 6       \\
13             & The First Affiliated Hospital of Xi an Jiaotong University,Xian Central Hospital                                                                                          & 25               & 17      \\
14             & The Second Xiangya Hospital of Central South University                                                                                                                   & 64               & 32      \\
15             & \begin{tabular}[c]{@{}l@{}}Zhongda Hospital, School of Medicine, Southeast University\end{tabular}                         & 50               & 50      \\
16             & Huaxi MR Research Center, West China Hospital of Sichuan University                                                                                                       & 31               & 31      \\
17             & Department of Psychiatry,The First Affiliated Hospital of Chongqing Medical University                                                                                    & 47               & 44      \\
18             & \begin{tabular}[c]{@{}l@{}}The First Affiliated Hospital, College of Medicine, Zhejiang University\end{tabular}                                 & 21               & 20      \\
19             & Anhui Medical University                                                                                                                                                  & 51               & 36      \\
20             & Faculty of Psychology, Southwest University                                                                                                                               & 282              & 251     \\
21             & Beijing Anding Hospital, Capital Medical University                                                                                                                       & 86               & 70      \\
22             & The Institute of Mental Health,Second Xiangya Hospital of Central South University                                                                                        & 30               & 20      \\
23             & Mental Health Center, West China Hospital, Sichuan University                                                                                                             & 32               & 30      \\
24             & First Affiliated Hospital of Kunming Medical University                                                                                                                   & 32               & 31      \\
25             & Department of Neurology,Affiliated ZhongDa Hospital of Southeast University                                                                                               & 89               & 63      \\ \bottomrule \hline
                     
\end{tabular}
\end{sidewaystable}

\begin{sidewaystable}[hbt!]
\renewcommand{\thetable}{S2}
\centering
\caption{Data acquisition parameters of selected sites.}
\label{Data acquisition parameters of selected sites}
\begin{tabular}{@{}lllllllllll@{}}
\hline
\toprule
Serial number & Scanner                       & Coil & TR (ms) & TE (ms) & Flip angle & Thickness/gap  & Slice number & Timepoints & Voxel size     & FOV     \\ \midrule \hline
1              & Siemens Tim Trio 3T           & 32   & 2000    & 30      & 90         & 4.0 mm/0.8 mm  & 30           & 210        & 3.28*3.28*4.80 & 210*210 \\
2              & Philips Achieva 3T            & 8    & 2000    & 30      & 90         & 4.0 mm/0 mm    & 37           & 200        & 1.67*1.67*4.00 & 240*240 \\
3              & Siemens 1.5 T                 & 16   & 2000    & 40      & 90         & 5.0mm/1.25mm   & 26           & 150        & 3.75*3.75*6.25 & 240*240 \\
5              & GE Signa 3T                   & 32   & 3000    & 30      & 90         & 5.0mm/0 mm     & 22           & 100        & 3.75*3.75*5.00 & 240*240 \\
6              & Siemens Tim Trio 3T           & 32   & 2000    & 30      & 70         & 4mm/0mm        & 33           & 180        & 3.59*3.59*4.00 & 230*230 \\
7              & GE discovery MR750            & 8    & 2000    & 30      & 90         & 3.2 mm/0 mm    & 37           & 184        & 2.29*2.29*3.20 & 220*220 \\
8              & GE Signa 3T                   & 8    & 2000    & 30      & 90         & 3.0 mm/0 mm    & 35           & 200        & 3.75*3.75*3.00 & 240*240 \\
9              & GE Discovery MR750 3.0T       & 8    & 2000    & 25      & 90         & 3.0 mm/1.0 mm  & 35           & 200        & 3.75*3.75*4.00 & 240*240 \\
10             & Siemens Tim Trio 3T           & 32   & 2000    & 30      & 90         & 3.0 mm/1.52 mm & 32           & 212        & 3.75*3.75*4.52 & 240*240 \\
11             & GE Signa 3T                   & 8    & 2000    & 30      & 90         & 5 mm           & 33           & 200        & 3.75*3.75*5.00 & 240*240 \\
12             & GE Signa 3T                   & 8    & 2000    & 30      & 90         & 5 mm           & 33           & 240        & 3.75*3.75*4.00 & 240*240 \\
13             & GE Excite 1.5T                & 16   & 2500    & 35      & 90         & 4 mm/0 mm      & 36           & 150        & 4.00*4.00*4.00 & 256*256 \\
14             & Siemens Tim Trio 3T           & 32   & 2500    & 25      & 90         & 3.5 mm/0 mm    & 39           & 200        & 3.75*3.75*3.50 & 240*240 \\
15             & Siemens Verio 3.0T MRI        & 12   & 2000    & 25      & 90         & 4 mm/0 mm      & 36           & 240        & 3.75*3.75*4.00 & 240*240 \\
16             & GE Signa 3T                   & 8    & 2000    & 30      & 90         & 5mm/0mm        & 30           & 200        & 3.75*3.75*5.00 & 240*240 \\
17             & GE Signa 3T                   & 8    & 2000    & 40      & 90         & 4.0 mm/0 mm    & 33           & 240        & 3.75*3.75*4.00 & 240*240 \\
18             & Philips Achieva 3.0 T scanner & 8    & 2000    & 35      & 90         & 5.0/1.0 mm     & 24           & 200        & 1.67*1.67*6.00 & 240*240 \\
19             & GE Signa 3T                   & 8    & 2000    & 22.5    & 30         & 4.0 mm/0.6 mm  & 33           & 240        & 3.44*3.44*4.60 & 220*220 \\
20             & Siemens Tim Trio 3T           & 12   & 2000    & 30      & 90         & 3.0 mm/1.0 mm  & 32           & 242        & 3.44*3.44*4.00 & 220*220 \\
21             & Siemens Tim Trio 3T           & 32   & 2000    & 30      & 90         & 3.5 mm/0.7 mm  & 33           & 240        & 3.12*3.12*4.20 & 200*200 \\
22             & Philips Gyroscan Achieva 3.0T & 32   & 2000    & 30      & 90         & 4.0 mm/0 mm    & 36           & 250        & 1.67*1.67*4.00 & 240*240 \\
23             & Philips Achieva 3.0T TX       & 8    & 2000    & 30      & 90         & 4.0 mm/0 mm    & 38           & 240        & 3.75*3.75*4.00 & 240*240 \\
24             & GE Signa 1.5T                 & 8    & 2000    & 40      & 90         & 5/1mm          & 24           & 160        & 3.75*3.75*6.00 & 240*240 \\
25             & Siemens Verio 3T              & 12   & 2000    & 25      & 90         & 4.0mm/0 mm     & 36           & 240        & 3.75*3.75*4.00 & 240*240 \\ \bottomrule \hline
\end{tabular}
\end{sidewaystable}

\begin{table}[hbt!]
\renewcommand{\thetable}{S3}
\centering
\caption{Comparison of brain-PAD in different models.}
\label{Comparison of brain-PAD in different models}
\begin{tabular}{@{}lccc@{}}
\hline
\toprule
Model      & MDD-NC(Mean-PAD) & $P$ value  & Cohen's \textit{d} \\ \midrule \hline
SVM-linear & 5.215            & 4.09e-09 & 0.367              \\
EN         & 3.282            & 5.88e-05 & 0.255              \\
Ridge      & 5.186            & 5.22e-09 & 0.364              \\
BR         & 4.532            & 3.55e-08 & 0.348              \\ \bottomrule \hline
\end{tabular}
\end{table}

\begin{table}[hbt!]
\renewcommand{\thetable}{S4}
\centering
\caption{Performance of different models.}
\label{Performance of different models}
\begin{tabular}{lccc}
\hline
\multicolumn{4}{c}{Validation set}                                 \\ \hline
Model        & MAE            & MSE               & R$^2$             \\
SVM-linear   & 8.9308 $\pm$ 0.6309  & 132.4733 $\pm$ 20.3135  & 0.4482 $\pm$ 0.0771  \\
SVM-RBF      & 11.6761 $\pm$ 0.8564 & 231.1530 $\pm$ 30.6658  & 0.04127 $\pm$ 0.0647 \\
RandomForest & 15.1180 $\pm$ 1.3921 & 433.7388 $\pm$ 57.8672  & -0.8006 $\pm$ 0.1541 \\
XGBoost      & 23.3025 $\pm$ 1.8482 & 776.5373 $\pm$ 110.7799 & -2.2516 $\pm$ 0.5404 \\ \hline
\multicolumn{4}{c}{Test set}                                       \\ \hline
SVM-linear   & 9.6537 $\pm$ 0.3379  & 148.42189 $\pm$ 9.4821  & 0.3051 $\pm$ 0.0443  \\
SVM-RBF      & 11.0753 $\pm$ 0.0464 & 205.1567 $\pm$ 3.4306   & 0.0395 $\pm$ 0.0160  \\
RandomForest & 15.4476 $\pm$ 0.1794 & 408.0862 $\pm$ 8.6558   & -0.9104 $\pm$ 0.0405 \\
XGBoost      & 34.2420 $\pm$ 0.1923 & 1590.5002 $\pm$ 22.9607 & -6.4457 $\pm$ 0.1074 \\ \hline
             &                &                   &               
\end{tabular}
\end{table}

\begin{table}[hbt!]
\renewcommand{\thetable}{S5}
\centering
\caption{Comparison of brain-PAD between different clinical characteristics.}
\label{Comparison of brain-PAD between different clinical characteristics}
\begin{tabular}{lcccccc}
\hline
\toprule
MDD vs NC         & Number & Brain-PAD & $P$ value  & SD    & Cohen's $d$ & 95\% CI   \\ \midrule \hline
MDD               & 1276   & 4.43                 & 3.49e-08 & 15    & 0.35      & 1.86-3.91 \\
First-episode MDD & 538    & 4.19                 & 7.80e-06 & 14.99 & 0.33      & 1.63-4.83 \\
Recurrent episode & 282    & 2.56                 & 2.10e-02 & 14.85 & 0.2       & -4.51     \\
antidepressant users     & 408    & 4.75                 & 3.99e-06 & 15.62 & 0.36      & 0.91-4.65 \\
Medication free   & 447    & 2.66                 & 7.16e-03 & 15.41 & 0.21      & -3.43     \\
Male              & 463    & 5.07                 & 1.85e-07 & 14.89 & 0.4       & 1.60-4.83 \\
Female            & 813    & 4.06                 & 2.30e-06 & 15.06 & 0.32      & 1.37-3.90 \\ \bottomrule \hline
\end{tabular}
\end{table}

\begin{table}[hbt!]
\renewcommand{\thetable}{S6}
\centering
\caption{Comparison of brain-PAD in education years and illness duration months.}
\label{Comparison of brain-PAD in education years and illness duration months}
\begin{tabular}{lcc}
\hline
\toprule
MDD subgroup                          & Number & Brain-PAD \\ \midrule \hline
Education\textless{}12                & 634    & 4.2                  \\
Education\textgreater{}=12            & 642    & 1.92                 \\
Duration \textless{}6                 & 329    & 4.05                 \\
Duration \textless{}12                & 461    & 3.7                  \\
6\textless{}= Duration \textless{}12  & 132    & 2.83                 \\
12\textless{}= Duration \textless{}24 & 130    & 2.48                 \\
Duration \textgreater{}=12            & 524    & 2.01                 \\
Duration \textgreater{}=24            & 394    & 1.86                 \\ \bottomrule \hline      
\end{tabular}
\end{table}

\end{document}